\documentclass[11pt,a4paper]{article}
\usepackage{jheppub}
\usepackage{graphicx,amsmath}
\usepackage{amsmath, amsfonts, amsthm, amssymb, graphicx}
\newcommand{\be}{\begin{equation}}
\newcommand{\ee}{\end{equation}}
\newcommand{\ba}{\begin{eqnarray}}
\newcommand{\ea}{\end{eqnarray}}
\newcommand{\del}{\partial}
 at 10truept

\newcommand{\beq}{\begin{equation}}
\newcommand{\eeq}{\end{equation}}
\newcommand{\bea}{\begin{eqnarray}}
\newcommand{\eea}{\end{eqnarray}}
\newcommand{\beqa}{\begin{eqnarray}}
\newcommand{\eeqa}{\end{eqnarray}}

\title{A stringy  perspective on the coincidence problem}
\author{Francesc Cunillera }

\author{Antonio Padilla} 

\affiliation{School of Physics and Astronomy, 
University of Nottingham, Nottingham NG7 2RD, UK} 

\emailAdd{francesc.cunilleragarcia@nottingham.ac.uk}
\emailAdd{antonio.padilla@nottingham.ac.uk}

\abstract{We argue that, for string compactifications broadly consistent with swampland constraints, dark energy is likely to signal the beginning of the end of our universe as we know it, perhaps even through decompactification,  with possible  implications for  the cosmological coincidence problem. Thanks to the scarcity (absence?) of stable de Sitter vacua, dark energy in string theory  is assumed to take the form of a quintessence field in slow roll.  As it rolls, a tower of heavy states will generically descend, triggering an apocalyptic phase transition in  the low energy cosmological dynamics after at most a few hundred Hubble times. As a result, dark energy domination cannot continue indefinitely and there is at least  a percentage chance that we find  ourselves in the first Hubble epoch. We use a toy model of quintessence coupled to a tower of heavy states to explicitly demonstrate the breakdown in the cosmological dynamics as the tower becomes light. This occurs through a large number of corresponding particles being produced after a certain time, overwhelming quintessence.  We also discuss some implications for early universe inflation. }

\date{\today}
\begin{document} 
\maketitle


\section{Introduction} \label{sec:intro}
A slew of observations, from supernova \cite{SN1,SN2} to the cosmic microwave background \cite{CMB},  point to a standard model of cosmology in which the universe  was dominated by radiation at early times, then by matter, before entering the current phase of dark energy domination.  The microscopic origin of dark energy has not been established  although on cosmological scales we know that it behaves like a fluid whose equation of state is very close to $\omega_{DE} \approx -1$ \cite{derev}.  This leads to a near constant energy density, consistent with a cosmological constant \cite{dS} or a slowly rolling quintessence field \cite{quin1a,quin1b,quin2}.  Assuming the underlying physics is governed by quantum field theory, the density of dark energy should be boosted by the energy of the quantum vacuum to values far in excess of those we observe, a puzzle normally referred to as  the cosmological constant problem  (for reviews see \cite{wein,pol,cliff, me}).  A related but qualitatively distinct puzzle is the {\it cosmological coincidence problem}. This  is usually presented as a simple question: why now? \cite{stein,Zlatev} Why do we happen to live at a time when the energy density of matter and dark energy are roughly comparable?  It is this latter problem that we seek to discuss in this paper, in the context of string compactifications and a consistent theory of quantum gravity. 

Let us unpack the problem in a little more detail. The total age of the universe is estimated to be around 13.7 Gyrs \cite{CMB}. The early phase of radiation domination represents only a small fraction of that history, giving way to matter after around 51 Kyrs, at a redshift of  $z_{eq} \approx 3400$. The transition to dark energy domination  occurred much later, after around 3.5 Gyrs, at a redshift of $z_{de} \approx 0.5$. This represents roughly a quarter of the universe's current lifetime \cite{coinrev}.  This means that the universe has been matter dominated for a significant fraction of its history, so much so that its density today is still comparable to that of dark energy.

Why is this a problem? The issue is the rate at which dark energy dilutes in comparison to matter, if indeed it dilutes at all. If dark energy is a cosmological constant, as in the standard $\Lambda$CDM cosmology,  its energy density remains constant while the energy density of matter falls off exponentially quickly once we enter the accelerating phase.  If we take the standard scenario at face value, dark energy will dominate indefinitely. It seem implausible that we should find ourselves so close to the dawn of dark energy domination, within a single Hubble time.  By way of comparison, we note that planetary orbits are expected to exist for another $10^5$ Hubble times \cite{future}. 

There have been several proposals that touch upon aspects of this problem (see \cite{coinrev}).  These range from  anthropic considerations \cite{anth1,anth2,anth3,anth4} to   k-essence scenarios where dark energy is triggered by matter-radiation equality, coming to dominate within a few billion years  \cite{track1, track2, ed}.   Although the latter go some way to explaining the (limited) duration of the matter dominated epoch, they do not address the main question of why we happen to find ourselves so close to the dawn of dark energy domination. 

This particular question can be addressed  in apocalyptic fashion by bringing the universe to a rapid conclusion.  The basic idea is that dark energy domination begins with acceleration before triggering cosmic Armageddon, within a few efolds. The solution to the coincidence problem follows because the accelerated epoch is cut short - it does not go on indefinitely and endures for a similar time to matter domination. This was shown to occur in phantom cosmologies \cite{phantom,anth4}, linear quintessence \cite{lin,anth4} and sequestering scenarios \cite{seq}.  The purpose of this paper is to investigate the extent to which  string theory may point to a similar resolution of the coincidence problem via the swampland conjectures \cite{swamp1,swamp2,distance1,swamp3,swamp4}. Our goal is to be as generic as possible  - some closely related ideas can, of course, be found in the swampland cosmology paper \cite{swampcos} and in  \cite{Scherrer}, in  the context of the Ijjas-Steinhardt cyclic models \cite{Ijjas}.

\section{Generic idea} \label{sec:idea}
We begin with the distance conjecture  \cite{distance1,distance2,distance3,distance4}, which is one of the most well studied  and least controversial of the swampland conjectures (see also \cite{moredist1,moredist2,moredist3,moredist4,moredist5,moredist6,moredist7,branes}). This states the following: consider two points in field space, $\phi_0$ and $\phi_0+\Delta \phi$,  separated by a geodesic distance $\Delta \phi$. As $\Delta \phi \to \infty$, there exists an infinite tower of states whose mass become exponentially light, 
\be
m(\phi_0+\Delta \phi) \sim m(\phi_0)e^{-\beta \frac{|\Delta \phi| }{M_{Pl}}} \label{distcon}
\ee
for some positive constant,  $\beta$, that we typically expect to be $\mathcal{O}(1)$. The offending tower of states is often associated with Kaluza-Klein modes or winding modes, depending on the direction of motion in moduli space.  For this reason, we take the initial mass $m(\phi_0)$ to be given by the scale of compactification, $1/R$, which  could be as low as a few meV in a braneworld setting where Standard Model fields are confined to a $3$-brane, although generically we expect it to be much larger, perhaps even just short of the Planck scale, $M_{Pl} \sim 10^{18}$ GeV. 

In its refined form, the de Sitter swampland conjecture  \cite{dscon1,chet,dscon2} concerns the form of  the potential $V(\phi)$ for scalar fields  in a low energy effective theory.  Assuming the effective theory  is obtained from a consistent theory of quantum gravity, the potential must satisfy either
\be
|\nabla V | \geq\frac{c}{M_{Pl}} V  \label{ds1}
\ee
or
\be
\text{min} \left( \nabla_i \nabla_j V\right) \leq -\frac{c'}{M_{Pl}^2} V \label{ds2}
\ee
where $c, c'$ are universal positive constants of $\mathcal{O}(1)$ and $\text{min} \left( \nabla_i \nabla_j V\right) $ is the minimum eigenvalue of the corresponding Hessian. Note that the conjecture forbids the existence of stable de Sitter vacua in string theory, for which we would have to have $V>0$. It does, however, allow for  de Sitter vacua with a tachyonic instability of order the corresponding Hubble time $H^{-1} \sim M_{Pl}/\sqrt{V}$. Some constraints on the scale of the tachyon were recently derived in the context of 10D supergravity \cite{and}.

The absence of stable de Sitter vacua points towards a dynamical model of dark energy.  Of course, building a reliable model of quintessence within string theory is not without its own challenges \cite{Heb}. Nevertheless, we begin with a model of quintessence as a canonical scalar field, $\phi$, moving under the influence of a potential $V$.  Here we imagine that quintessence is generically described by the saxions of string theory with a non-compact field range. The energy density and pressure stored in the field are given respectively by $\rho_\phi=\frac12 \dot \phi^2+V$ and $p_\phi =\frac12 \dot \phi^2-V$. The dynamics of the scalar is governed by the following field equation
\be
\ddot \phi +3H\dot \phi+V'(\phi)=0
\ee
where $H(t)$ is the Hubble parameter and ``dot" denotes differentiation with respect to cosmological time.  Our goal is to argue that dark energy domination is relatively short lived on account of the motion of the field  towards the infinite points in moduli space. Any motion of the field prior to dark energy domination will only reduce the amount it is allowed to move afterwards, bringing dark energy to an even quicker conclusion. Therefore, the most conservative scenario is to assume negligible motion of the moduli fields  until dark energy finally begins to dominate.  This is, in any event, likely as the field will be held up by Hubble friction. 

Once the dark energy field has come to dominate, to be compatible with the observed equation of state, it must be in slow roll, $\frac12 \dot \phi^2 \ll V, ~|\ddot \phi | \ll 3H|\dot \phi| $. Furthermore, we may assume that $H \approx H_0 \sim 10^{-33}$ eV $\sim 10^{-60} M_{Pl}$.  With these approximations, consider the field excursion in a short time $\delta t$. This is given by 
\be
\delta \phi \approx -\frac{V'}{3H_0} \delta t
\ee
If we accept the refined de Sitter conjectures, then one of \eqref{ds1} or \eqref{ds2} must hold. Let's assume we satisfy the condition on the gradient, given by \eqref{ds1}. It now follows that 
\be
| \delta \phi | \gtrsim \frac{c}{M_{Pl}}   \frac{V}{3H_0} \delta t \sim \mathcal{O}(1)  M_{Pl} H_0 \delta t
\ee
where we have used the Friedmann equation during dark energy domination  $H_0^2 \approx \frac{\rho_\phi}{ 3 M_{Pl}^2} \approx  \frac{V}{ 3 M_{Pl}^2} $ and the fact that $c$ is assumed to be $\mathcal{O}(1)$. We immediately see that the dark energy field rolls roughly a Planck unit in a Hubble time.  We now consider the implications for the distance conjecture \eqref{distcon}, assuming the tower of new states are initially very heavy, with masses $m(\phi_0) \sim M_{Pl}$ close to the Planck scale. After a single Hubble time, the dark energy field will move by around one Planck unit.  This only corresponds to a fractional change in the masses in the tower, and certainly not enough to contaminate the low energy physics. 

How far is the field allowed to roll before we have to start worrying about it?  In the local neighbourhood of the Earth the field should not be displaced from $\phi_0$ by more than around 30 Planck units. Anything more than that would bring the mass of the tower down from Planck scale to the scale of collider physics, opening up the possibility of producing these states at the LHC.  Of course, the details of this depend on the nature of the coupling between the tower and the Standard Model fields. Furthermore,  none of these considerations are relevant on cosmological scales, where we can certainly allow the field to move much further.  To ensure that the tower remains decoupled from the low scale cosmological dynamics, we conservatively impose a maximum displacement of  $|\Delta \phi|_\text{max} \sim \frac{M_{Pl}}{\beta} \ln(M_{Pl}/H_0)$. For $\beta \sim \mathcal{O}(1)$ this corresponds to a displacement of around 140 Planck units. Assuming the field continues to roll a Planck unit in every Hubble time, we see that the tower of states  will trigger a transition in the cosmological dynamics after no more than $\mathcal{O}(100)$ Hubble times.  The coincidence problem isn't solved but  it is significantly ameliorated. If a generic model of dark energy is destined to last at most $\mathcal{O}(100)$ Hubble epochs, and all epochs are equally probable,  we might expect there to be at least a percentage chance that we find ourselves in the first epoch. By way of comparison, we note that when Leicester City won the premier league in 2016, they started the season as 5000-1 outsiders.  

What if we assume that we satisfy the second of the two de Sitter swampland criteria \eqref{ds2}, rather than the first \eqref{ds1}? If the condition on the gradient \eqref{ds1} is violated, but the condition on the Hessian  \eqref{ds2} holds the dark energy field may move considerably less than a Planck unit in a Hubble time.  In fact, if we imagined the field sitting in a region where the gradient of the potential is negligible, we might even imagine it staying there indefinitely, giving a neverending period of dark energy.  However, this conclusion is too quick.  Quantum fluctuations will guarantee a displacement in the field of at least $\mathcal{O}(H_0)$ in the first Hubble time.   This initial displacement will grow thanks to the fact that the Hessian condition implies a tachyonic mass for the fluctuations in the dark energy field,  $\mu^2 = V''(\phi)<0$ with $|\mu^2| \gtrsim  \frac{c'}{M_{Pl}^2} V  \approx  3c' H_0^2$.   For $c' \sim \mathcal{O}(1)$, the corresponding instability can be as slow as a Hubble time but even so, its effect is amplified over several Hubble times by exponential growth.  There are two possibilities: the first is that the instability triggers a rapid transition  which brings the acceleration to a premature end  (as desired for the coincidence problem).  This could occur, for example,  by the potential changing sign so that  we no longer have a quasi de Sitter expansion. Alternatively, the background cosmology could remain roughly unchanged, at least beyond a few Hubble times. If this is the case, the tachyonic instability  amplifies the initial displacement to a value\footnote{To leading order, the scalar satisfies an equation $\ddot \phi +3H_0 \dot \phi-|\mu^2| \phi=0$, where we recall that $\mu^2 = V''(\phi)<0$ with $|\mu^2| \gtrsim  \frac{c'}{M_{Pl}^2} V  \approx  3c' H_0^2$. The general solution is then given by a sum of exponentials $e^{\lambda_\pm t}$ where $\lambda_\pm=\frac{-3H_0 \pm \sqrt{9H_0^2+4 |\mu^2| }}{2}$, which includes a growing mode, with $\lambda_+ \gtrsim H_0$,  thanks to the tachyonic instability with $|\mu^2|  \gtrsim H_0^2$.}
\be \label{tach}
| \delta \phi | \gtrsim  H_0 e^{\mathcal{O}(1) H_0 \delta t}
\ee

In the latter scenario, the tower of massive states would remain decoupled from the cosmological dynamics until $| \delta \phi |  \approx |\Delta \phi|_\text{max} \sim \frac{M_{Pl}}{\beta} \ln(M_{Pl}/H_0)$, at which point a transition is inevitable. For $\beta \sim \mathcal{O}(1)$, this will occur within at most  $\ln[M_{Pl}/H_0 \ln(M_{Pl}/H_0))] \sim 143$ Hubble times, so our conclusions are unchanged, and the coincidence problem isn't as serious as we previously thought. 

Of course, the de Sitter conjecture is less well established than the distance conjecture, and, in the simplest scenarios, may even be at odds with local measurements of $H_0$ \cite{colg}. With this in mind, what can we say if  we  deny the validity of both criteria  \eqref{ds1} and \eqref{ds2} and abandon the de Sitter conjecture altogether? In this instance, we cannot rule out the possibility that the current phase of acceleration is approaching a stable de Sitter configuration in which  dark energy continues for an exponentially large number of Hubble epochs. If this is the case, then the coincidence problem is as problematic as ever.  However, we might tentatively speculate that the de Sitter conjecture is really a statement about what is {\it generic} within consistent models of dark energy within string theory.  Of course, it is much too early to make any definitive statement in this regard. Nevertheless, if it happens to be true that the generic scenarios are those for which one of the criteria \eqref{ds1} or \eqref{ds2} hold,  our results go through and the coincidence problem is tamed.  We might also worry about the fact we have assumed dark energy to be a single canonical scalar.   However, we  expect this to capture the generic dynamics of fields moving through moduli space,  with our canonical scalar tangential to the trajectory and all the orthogonal directions stabilised.

\section{A toy model} \label{sec:disc}
To better understand how an accumulation of light states can impact the cosmological evolution at late times, we consider a toy model of dark energy described by the following Lagrangian 
\be
\mathcal{L}=-\frac12 (\partial \phi)^2-\mu^4e^{- \frac{\alpha \phi}{M_{Pl}}} 
+\sum_{n=1}^{\infty}-\frac12 (\del \varphi_n)^2
-\frac12 n^2 M_{KK}^2 e^{-2  \frac{\beta \phi}{M_{Pl}}} \varphi_n^2 
\ee
Here $\phi$ is the quintessence field, driving dark energy, taken to be in slow roll on a runaway potential of the form $\mu^4e^{-\frac{\alpha \phi}{M_{Pl}}}$ where $\mu^4 = 3M_{pl }^2H_0^2$ is the  scale  of dark energy and $\alpha$ is some order one positive number.  To be compatible with observations we require $\alpha<0.6$ \cite{swampcos}, which is not in conflict with the swampland constraints on the potential \eqref{ds1} and \eqref{ds2}. In addition we have a tower of heavy states, $\varphi_n$, whose masses are originally set by some high scale $M_{KK}$ (imagined to be the Kaluza-Klein scale associated with the compact  internal manifold), becoming exponentially light as the dark energy field, $\phi$, rolls off to infinity. The rate at which the tower becomes light is set by the coupling $\beta$, which is also assumed to be order one and positive. 

At the  dawn of dark energy domination, at time $t_0$, we assume that the quintessence field is far away from the  tails of the exponentials, $\phi \sim \phi_0 \ll M_{Pl}/\alpha, M_{Pl}/\beta$.  Since the effective mass of the Kaluza-Klein tower, $M_{KK} e^{-  \frac{\beta \phi_0}{M_{Pl}}} \sim M_{KK} \gg H_0$ is high in this regime, far above the scale of the cosmological evolution, the Kaluza-Klein states are decoupled from the dynamics.  As a result, the quintessence field satisfies the following classical equation of motion on a homogenous and isotropic background
\be
\ddot \phi+3H\dot \phi-\frac{\alpha}{M_{Pl}}\mu^4e^{- \frac{\alpha \phi}{M_{Pl}}}=0
\ee
For a quasi-de Sitter expansion, with $H\approx H_0$, we find that $\phi \approx \phi_0+ \lambda (t-t_0)$ where 
\be
\lambda= \frac{\alpha \mu^4}{3 M_{Pl}H_0} =  \alpha M_{Pl} H_0< M_{Pl} H_0
\ee
This approximation works well as long as $\Delta \phi < M_{Pl}/\alpha$, or equivalently, $\Delta t <H_0^{-1}/\alpha^2$.  Of course, acceleration will continue beyond this time, only at a lower scale. There is a wealth of literature on the dynamics of similar quintessence models (for a review, see \cite{derev}). In this letter, we wish to briefly explore another effect that is far less well studied - the time dependence on the mass of the Kaluza-Klein tower as the field begins to slowly roll. On the quasi-de Sitter background with constant curvature, this is given by
\be
M_{KK}^\text{eff}(t)=M_{KK} e^{-  \frac{\beta s}{M_{Pl}}} \approx M_{KK}  e^{-  \epsilon (t-t_0)} 
\ee
where $\epsilon=\beta \lambda/M_{Pl} =\alpha\beta H_0$. This will drive particle production in the Kaluza-Klein sector, kicking in as soon as the states stop being decoupled,  $M_{KK}^\text{eff}(t) \lesssim H_0$.  A complete analysis of this phenomena requires a detailed numerical study of particle creation on the dynamical background, taking into account the effect of the time varying mass and the de Sitter cosmology. This work \cite{dsstuff} is now underway but beyond the scope of the current letter.  To get some immediate insight into what might happen we neglect the  curvature of the background spacetime and focus on the particle production due  to a mass varying exponentially with time on a Minkowski geometry.  Crucially,  these approximations allow us to make {\it preliminary} analytic estimates but we should also acknowledge their limitations. In particular, we are implicitly assuming that the dark energy field, which feeds into the effective mass of the Kaluza Klein tower, continues to evolve linearly in time beyond the first Hubble epoch. In truth, the Hubble scale changes and the dark energy field picks up additional temporal dependence which may affect some of the details.  Also, we are neglecting the effect of spacetime curvature. This is less of an issue, as our interest here is  on the effect of the  changing mass of the Kaluza-Klein tower, as opposed to the effect of quantum fields propagating on de Sitter.  We also expect this approximation to accurately capture the physics on sub-horizon scales. 

With these caveats in mind, let us take a closer look at the $n$th state in the Kaluza-Klein tower on a Minkowski background, whose dynamics described by the following Lagrangian
\be
\mathcal{L}=-\frac12 (\del \varphi_n)^2
-\frac12 n^2 M_{KK}^2 e^{- 2 \epsilon t}  \varphi_n^2 
\ee
where we have also set $t_0=0$ (without loss of generality). Our analysis follows the standard techniques reviewed in detail in \cite{mukbook}, whose conventions we also follow.  As explained  in \cite{mukbook},  the state operator for the quantum field  can be expanded in terms of creation and annihilation operators, $\hat a_{\bf k}^\dagger$ and  $\hat a_{\bf k}$, in the usual way,
\be
\hat \varphi_n(t, {\bf x})=\frac{1}{\sqrt{2} }\int \frac{d^3 \bf  k}{(2\pi)^\frac32} \left[ e^{i\bf{k \cdot x}} \bar u_k(t) \hat a_{\bf k}+ e^{-i\bf{k \cdot x}  }u_k(t) \hat a^\dagger_{\bf k} \right]
\ee
where ``bar" denote the complex conjugate.  The mode functions are governed by  the equation for a time dependent  harmonic oscillator 
\be
\ddot u_{k}+\omega^2_k(t) u_k=0, \qquad \omega^2_k(t)=k^2+n^2 [M_{KK}^\text{eff}(t)]^2
\ee
and have  solutions that are conveniently expressed in terms of Hankel functions (of the first kind)
\be
u_k(t)=A H^{(1)}_{i\mu}(z)+B \bar H^{(1)}_{i\mu}(z) \label{uk}
\ee
where $z=\frac{n M_{KK} }{\epsilon} e^{-  \epsilon  t}$ and $\mu=\frac{k}{\epsilon}$.  Since $[M_{KK}^\text{eff}(t)]^2>0$, at any given time $t_*$ we can define the instantaneous vacuum state $|0 \rangle_*$ as the lowest energy state of the Hamiltonian at that time.  The mode functions that determine this state  satisfy the boundary condition \cite{mukbook}
\be
u_k(t_*)=\frac{1}{\sqrt{\omega_k(t_*)}}, \quad \dot u_k(t_*)=i \sqrt{\omega_k(t_*)}
\ee
This fixes the constants in \eqref{uk} so that
\begin{eqnarray}
A&=&\frac{\pi i e^{-\pi\mu} }{4 \sqrt{\omega_k(t_*)}}\left[z_*\bar H^{(1)}_{i\mu}{}'(z_*)+ \frac{\omega_k(t_*)}{\epsilon} i \bar H^{(1)}_{i\mu}(z_*)\right] \\
B&=&-\frac{\pi i e^{-\pi\mu} }{4 \sqrt{\omega_k(t_*)}}\left[z_* H^{(1)}_{i\mu}{}'(z_*)+ \frac{\omega_k(t_*)}{\epsilon} i  H^{(1)}_{i\mu}(z_*)\right]~
\end{eqnarray}
The mode functions that determine the ``in" vacuum at the beginning of the dark energy era, $u_k^\text{in}(t)$,  are given by these expressions for $A, ~B$ with the choice $t_*=0$. The mode functions  $u_k^\text{out}(t)$, that determine the ``out" vacuum at some later time, $T$,  are given by the same formulae but with $t_*=T$. As usual, the two can be related by  a Bogoliubov transformation, of the form $u_k^\text{in}(t)=\alpha_k (T) u_k^\text{out}(t)+\beta_k(T) \bar u_k^\text{out}(t)$.  At time $T>0$, the true vacuum state differs from the initial vacuum state, and so the latter contains particles. As explained in \cite{mukbook}, the mean particle number density for modes of momentum $\bf k$ is $N_k(T)=|\beta_k(T)|^2$ with corresponding energy density  $E_k(T)=\omega_k(T)N_k(T)$.  When we calculate this explicitly, it turns out that 
\be
N_k(T)=\frac{\pi^2}{16} e^{-2\pi\mu} \epsilon^2\left[X^2+Y^2\right]
\ee
where 
\be
X= \text{Im} \left[ \frac{ z_\text{in} H^{(1)}_{i\mu}{}'(z_\text{in} )}{\sqrt{\omega_k(0)}}\frac{ z_\text{out} \bar H^{(1)}_{i\mu}{}'(z_\text{out} )}{\sqrt{\omega_k(T)}} \right.  \left. - \frac{ \sqrt{\omega_k(0)} H^{(1)}_{i\mu}(z_\text{in} )}{\epsilon}\frac{\sqrt{\omega_k(T)} \bar H^{(1)}_{i\mu}(z_\text{out} )}{\epsilon} \right]
\ee
and
\be
Y= \text{Im} \left[ \frac{ z_\text{in} H^{(1)}_{i\mu}{}'(z_\text{in} )}{\sqrt{\omega_k(0)}} \frac{\sqrt{\omega_k(T)} \bar H^{(1)}_{i\mu}(z_\text{out} )}{\epsilon} \right.  \left. - \frac{ \sqrt{\omega_k(0)} \bar H^{(1)}_{i\mu}(z_\text{in} )}{\epsilon} \frac{ z_\text{out} \bar H^{(1)}_{i\mu}{}'(z_\text{out} )}{\sqrt{\omega_k(T)}}\right] 
\ee
Here $z_\text{in}=\frac{nM_{KK}}{\epsilon}$ and  $z_\text{out}=\frac{nM_{KK}}{\epsilon}e^{-\epsilon T}$.  We can obtain estimates for the energies stored in different momentum modes by using the approximations for Bessel and Hankel functions given in \cite{AS}. In particular,  we note that for large $ z \gg \mu^2+1$,  we have the following asymptotic expansion 
\be \label{approx1}
 H^{(1)}_{i\mu}(z) \approx
 \sqrt{\frac{2}{\pi z}} e^{\frac{\mu\pi}{2}+i\left(z-\frac{\pi}{4}\right)}\left[1+\frac{4 \mu^2+1}{8iz}+\ldots \right]
\ee
To obtain an approximation in the opposite limit,  for small $z$, we recall that $H^{(1)}_{i\mu}(z)=\frac{e^{\mu \pi} J_{i\mu} (z)-J_{-i\mu}(z)}{\sinh\mu\pi}$ and use the fact that 
\be \label{approx2}
J_{i\mu} (z) \approx \frac{e^{i\mu \ln \frac{z}{2} }}{ \Gamma(1+i\mu)}\left[1-\frac{z^2}{4(1+i\mu)}+\frac{z^4}{32(1+i\mu)i\mu}+\ldots\right]
\ee
whenever   $0<z \ll \sqrt{|\mu|+1} $.

\begin{figure}[h]
\includegraphics[width=\textwidth]{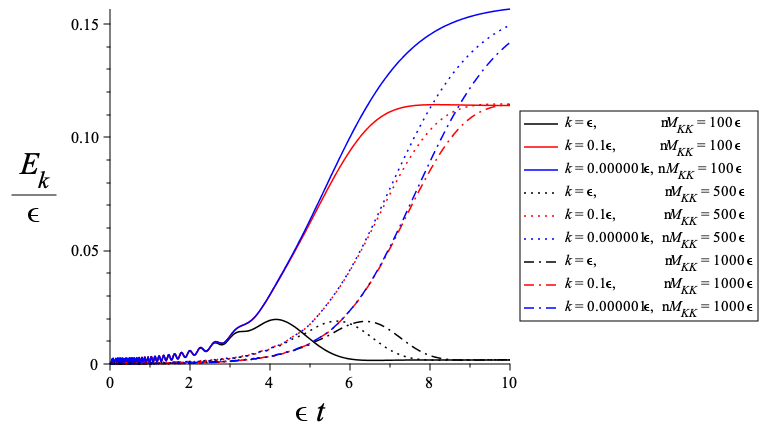}
\caption{Plot of energy versus time for  a range of scalar modes of different momentum  and  different values for the initial mass. These particular  plots were   produced using Maple 2020 and a numerical value of $\epsilon=0.01$. }\label{fig1}
\end{figure}

For high momentum modes, with $k \to \infty$, we find that
 $E_k \sim \frac{n^4 M_{KK}^4  \epsilon^2 }{4k^5}\left[\cosh^2\epsilon t-\cos^2kt\right]e^{-2\epsilon t}$. 
As expected, these modes are suppressed, being insensitive to the change in the mass of the field.  For modes of lower momentum, it is instructive to display the changes in the energy stored in each mode in a characteristic plot, such as the one shown in FIG. \ref{fig1}.  
The plot  shows the energy profile for modes running over a range of different momenta. We also vary over the initial mass of the scalar, $nM_{KK}$, or equivalently, over the level in the Kaluza-Klein tower.  For the $\varphi_n$ particles, at level $n$,  we see that the total energy density at late times is dominated by modes of momentum $k\lesssim \epsilon$, with significant particle production kicking in at a time $t_n \sim \frac{1}{\epsilon}\ln \left(\frac{nM_{KK}}{\epsilon}\right)$.  Using \eqref{approx1} and \eqref{approx2}, we can estimate  the energy stored in the low energy modes analytically, for $t \gg t_n$.  Since $nM_{KK} e^{-\epsilon t}  \ll \epsilon \ll nM_{KK}$ and $k \lesssim \epsilon$, we have $z_\text{out} \ll 1 \ll z_\text{in}$ and  $\mu \lesssim 1$, so that the energies approximate as
\be
E_{k} \approx 
\frac{k e^{-\frac{k\pi}{\epsilon}}}{2 \sinh\left(\frac{k\pi}{\epsilon}\right)}
\ee
In deriving this, we have used the relation $|\Gamma(i\mu)|^2 =\pi/\mu \sinh \mu\pi$, which follows from Euler's reflection formula and the fact that $\Gamma(-i\mu)=\overline{\Gamma(i\mu)}$. To obtain an estimate for the total  energy density stored in $\varphi_n$ particles at late times ($t \gg t_n$), we simply  integrate this result over all momentum. The result is
\be
\rho_n =\int d^3 {\bf k} E_k\approx \frac{\pi}{60}\epsilon^4
\ee
By itself, this would only have a tiny effect on the cosmological background, since $\epsilon^4 =(\alpha\beta)^4 H_0^4$, far below the scale of the critical density of the universe during the dark energy era, $M_{Pl}^2 H_0^2$. Of course, the formula will receive additional  corrections from the fact that the fields are actually propagating on a dynamical cosmological background, as opposed to Minkowski, although these are likely to be similarly suppressed, especially if $\alpha\beta \gtrsim 1$. Nevertheless,  by the time we have reached time $t_N \sim \frac{1}{\epsilon}\ln \left(\frac{NM_{KK}}{\epsilon}\right)$ for some large $N$, we have started to produce particles for each of the first $N$ levels in the Kaluza-Klein tower.  The {\it total} energy density of all these particles, in this approximation,  is given by
\be
\rho_\text{total} (t \sim t_N)=\sum_{n=1}^N \rho_n \approx \frac{N\pi}{60}\epsilon^4
\ee
This will start to affect the cosmological dynamics for $N\sim N_\text{crit}$ where $ N_\text{crit} \sim \frac{M_{Pl}^2}{\epsilon^2}$, at a time $t_\text{crit} \sim \frac{1}{\epsilon}\ln \left(\frac{N_\text{crit}M_{KK}}{\epsilon}\right)$.  If we assume $\alpha\beta \sim \mathcal{O}(1)$, then $\epsilon \sim H_0$ and so $N_\text{crit}\sim 10^{120}$. Assuming $M_{KK} \lesssim M_{Pl}$, so that $N_\text{crit}  \gg\sqrt{N_\text{crit}}  \gtrsim \frac{M_{KK}}{\epsilon}$, we see that  $t_\text{crit}\lesssim \mathcal{O}(400)/H_0$, confirming our expectation that the dark energy era will not last beyond a few hundred or so Hubble times, before the space begins to decompactify.   If we lower the underlying Kaluza-Klein scale, $M_{KK}$, or consider models with $\alpha\beta>1$, the dark energy era is cut short even earlier. 

By the time dark energy succumbs to the extra dimensions, a huge number of particles species have already entered the low energy effective theory. We might be worried that this creates a ``species problem" at some earlier time, lowering the scale at which gravity becomes strongly coupled. However, for $N<N_\text{crit}$ species, the scale of strong coupling scales as $\Lambda_{QG} \sim M_{Pl}/\sqrt{N}>H_0$. In  other words,  four-dimensional gravity does not become strongly coupled on cosmological scales prior to $t_\text{crit}$, at which point the effective four-dimensional description breaks down anyway. 

As emphasized earlier, everything we are saying is only relevant to cosmological dynamics. On shorter scales, for example in the lab or in the solar system, the  quintessence field may be displaced from its cosmological value, so much so that the Kaluza-Klein tower remains heavy and decoupled from the low energy physics.  Indeed, if quintessence is coupled to matter with gravitational strength, such a displacement is likely to be necessary in order to avoid fifth force constraints \cite{cham1,cham2, cham3}.  It would certainly be interesting to investigate the implications of this in more detail. 

 \section{Discussion} \label{sec:disc}
In this paper, we have argued that string theory compactifications consistent with swampland constraints automatically prevent the dark energy era from extending beyond a few hundred Hubble times. This alleviates the coincidence  problem to some degree, inasmuch as we now have as much as a percentage chance of finding ourselves in the first e-fold of acceleration.  Our conclusions draw on two key features of string compactifications: (i) the scarcity (absence?) of stable de Sitter vacua, suggesting that most dark energy models will be driven by a quintessence field in slow roll \cite{dscon1,dscon2}; (ii) the accumulation of a large number light states as the field rolls off towards infinity  \cite{distance1,distance2,distance3,distance4}. Generically, the dynamics is such that the descending  tower of light states induces a cosmological phase transition, bringing the dark energy era to a conclusion.  We were able to demonstrate this explicitly with a toy model, whereby particle creation in the Kaluza-Klein sector starts to overwhelm the cosmological background within a few hundred Hubble times.   This suggests we might even think of dark energy as opening the door to the decompactification of spacetime!  

From this stringy perspective, it seems that the  cosmological coincidence is readily reduced to one part in $\mathcal{O}(100)$ but can we do any better? Can we exploit the distance conjecture to bring the odds even further in? One speculative possibility is to consider dark energy a consequence of clockwork dynamics \cite{cwkde}. The clockwork construction allows for the existence of a naturally light dark energy field, in a theory with uniquely high scale couplings. But the set-up is brittle.  After the dark energy field has moved just a single Planck unit, we could imagine a tower of states contaminating the high scale physics and the spoiling the clockwork dynamics. Without the clockwork, the dark energy dynamics should also be spoilt. Whilst this idea is appealing we have not been able to identify a stringy realisation of the clockwork dark energy model, at least within the context of perturbative type IIA  and IIB supergravity \cite{otherpaper}.

Finally, it is natural to ask what would happen if we applied the same analysis to early universe inflation at a much higher scale than quintessence.  Because the Kaluza-Klein tower does not need to descend as far to contaminate the inflationary background, the inflaton excursion is limited to just a few Planck units. As a result, we would conclude that, generically, inflation should not extend beyond a few efolds. This seems problematic since inflation must last for at least 50 efolds in order  to address the horizon problem. Of course, these concerns aren't new - it is well known that early  universe inflation is in some tension with the swampland constraints \cite{swampcos}.   

An alternative approach could be to consider the axions as inflaton candidates, as opposed to the saxions described above. However, these models are also in tension with the swampland programme and in particular with the axionic weak gravity conjecture (aWGC) \cite{AHMNV}. The aWGC implies the existence of at least one instanton coupling electrically to an axion such that $f\times S_{\text{int}}\lesssim M_{Pl}$, where $f$ is the axion decay constant and $S_{\text{inst}}$ is the instanton action. If the instanton couples to the inflaton, which requires a super-Planckian field range, $S_{\text{inst}}<1$ and we lose parametric control of the low-energy theory.  We can get around this by assuming the existence  of heavy {\it spectator} axions that do not contribute to the inflationary dynamics and that satisfy the aWGC bound. However, stronger forms of the aWGC claim that {\it spectator} axions are not enough, imposing further hurdles on axion inflation models which remain unsolved \cite{Shiu1,Shiu2, Rudelius, Shiu3}.

Clearly inflation presents a different challenge to the string model builders than quintessence.  However, we can tentatively speculate that aspects of the swampland programme, and the de Sitter conjecture in particular, may conservatively be taken as  as an indication of what to expect from a generic  model of a scalar field in string theory, as opposed to a hard and fast rule. With this perspective, inflation would be cut short in a generic model,  but there may be  rare models allowing for many more efoldings. With anthropic reasoning, we can then argue that it was necessary for our universe to explore these rare models in order to grow large. For quintessence, the story is different. We cannot make the same anthropic arguments in favour of a rare model of quintessence that is long lived, so we are drawn instead to scenarios satisfying the constraints \eqref{ds1} and \eqref{ds2}. If quintessence scenarios satisfying these constraints can be found they will imply that the future of our universe  should not last for more than a hundred or so Hubble times.

\begin{acknowledgments}
A.P is  funded by an STFC Consolidated Grant and F.C by a University of Nottingham studentship. We would like to thank Michele Cicoli and Francisco G. Pedro for invaluable discussions. 
\end{acknowledgments}

\end{document}